\documentclass[conference]{IEEEtran}

\usepackage{tikz}
\usepackage{amsmath}
\usepackage{xcolor}
\usepackage{filecontents}

\usepackage{url}        % For \nolinkurl
\usepackage{hyperref}
\usepackage{booktabs}
\usepackage{amssymb}
\usepackage{amsfonts}

\usepackage{subfig}
\usepackage{wrapfig}
\usepackage{float}
\usepackage{multirow, graphicx}
\usepackage{xspace}

\usepackage{url}

\usepackage{sty/widetext} %show equation in two columns

\usepackage{stackrel}

\usepackage{pifont}% http://ctan.org/pkg/pifont
\newcommand{\cmark}{\ding{51}}%
\newcommand{\xmark}{\ding{55}}%

\usepackage{todonotes}
\usepackage{marginnote}
\newcommand{\ignore}[1]{}

\usepackage{enumitem}
\setlist[itemize]{leftmargin=*}

\usepackage{graphicx, multirow}

\usepackage{tabularx, multirow, rotating} %tables, dependent on multirow package.
\usepackage{subfig} %figures
\usepackage{algorithm} %algorithm
\usepackage{algpseudocode} %algorithm: use algorithmicx package, to refer to http://en.wikibooks.org/wiki/LaTeX/Algorithms_and_Pseudocode#Typesetting_using_the_algorithmicx_package
\usepackage{ulem} % strike out text
\usepackage{color} % colored text

\usepackage{listings} 
\lstdefinestyle{Oracle}{basicstyle=\ttfamily,
                        keywordstyle=\lstuppercase,
                        emphstyle=\itshape,
                        showstringspaces=true,
                        }
\makeatletter
\newcommand{\lstuppercase}{\uppercase\expandafter{\expandafter\lst@token
                           \expandafter{\the\lst@token}}}
\newcommand{\lstlowercase}{\lowercase\expandafter{\expandafter\lst@token
                           \expandafter{\the\lst@token}}}
\makeatother

\usepackage{tikz}

\newif\ifboldnumber

\algrenewcommand\alglinenumber[1]{%
  \footnotesize\ifboldnumber\bfseries\fi\global\boldnumberfalse#1:}

\usepackage{color, colortbl}
\definecolor{Gray}{gray}{0.9}
\definecolor{LightCyan}{rgb}{0.88,1,1}
\usepackage[first=0,last=9]{lcg}

\usepackage{colortbl}
\usetikzlibrary{calc}
\usepackage{zref-savepos}

\newcounter{NoTableEntry}
\renewcommand*{\theNoTableEntry}{NTE-\the\value{NoTableEntry}}

\usepackage{url}        % For \nolinkurl
\usepackage{hyperref}
\usepackage{diagbox} % add this in your preamble if not already included

\ifCLASSINFOpdf
\else
\fi
\begin{document}
\title{Position Paper: Denial-of-Service against Multi-Round Transaction Simulation}

\author{\IEEEauthorblockN{
Yuzhe~Tang\IEEEauthorrefmark{1},
Yibo~Wang\IEEEauthorrefmark{2},
Wanning~Ding\IEEEauthorrefmark{1},
Jiaqi~Chen\IEEEauthorrefmark{1},
Taesoo~Kim\IEEEauthorrefmark{3}
% <-this % stops a space}
}
\IEEEauthorblockA{\IEEEauthorrefmark{1}Syracuse University,
{\{ytang100, wding04, jchen217\}}@syr.edu}
\IEEEauthorblockA{\IEEEauthorrefmark{2}University of Wyoming,
ywang34@uwyo.edu
}
\IEEEauthorblockA{\IEEEauthorrefmark{3}Georgia Institute of Technology and Microsoft,
taesoo@gatech.edu
}
}


%Telephone: (800) 555--1212, Fax: (888) 555--1212}


\IEEEoverridecommandlockouts
\makeatletter\def\@IEEEpubidpullup{6.5\baselineskip}\makeatother

\maketitle

% \usepackage{tikz}
\newcommand*\rectangled[1]{%
  \tikz[baseline=(R.base)]\node[draw,rectangle,minimum width=1em,minimum height=1em,inner sep=0pt,align=center](R) {\strut #1};\!
}

\providecommand{\bbPreCk}{\textsc{p}\xspace}
\providecommand{\bbPreMem}{\textsc{m}\xspace}
\providecommand{\bbPostCk}{\textsc{q}\xspace}
\providecommand{\badAdvance}{\ifmmode\mathbb{S}\else$\mathbb{S}$\fi\xspace}
\providecommand{\badExhaust}{\ifmmode\mathbb{H}\else$\mathbb{H}$\fi\xspace}
\providecommand{\badExclude}{\ifmmode\mathbb{X}\else$\mathbb{X}$\fi\xspace}
\providecommand{\badInclude}{\ifmmode\mathbb{I}\else$\mathbb{I}$\fi\xspace}

\providecommand{\codescroll}{\textsc{s}\xspace}
\providecommand{\codezksync}{\textsc{z}\xspace}
\providecommand{\codeflashtwo}{\textsc{f}\xspace}
\providecommand{\codeflashthree}{\textsc{g}\xspace}
\providecommand{\codezkevm}{\textsc{v}\xspace}

\providecommand{\azero}{\textsc{A0}}
\providecommand{\aone}{\textsc{A1}}
\providecommand{\atwo}{\textsc{A2}}
\providecommand{\athree}{\textsc{A3}}
\providecommand{\afour}{\textsc{A4}}
\providecommand{\afive}{\textsc{A5}}

\providecommand{\mzero}{\textsc{M0}}
\providecommand{\mone}{\textsc{M1}}
\providecommand{\mtwo}{\textsc{M2}}
\providecommand{\mthree}{\textsc{M3}}
\providecommand{\mfour}{\textsc{M4}}
\providecommand{\mfive}{\textsc{M5}}

%%%% old version %%%%%
\providecommand{\preck}{{\sc \rectangled{p}\xspace}}
\providecommand{\postck}{{\sc \rectangled{q}\xspace}}

\providecommand{\exec}{{\sc \rectangled{e}\xspace}}
\providecommand{\execsuccess}{{\sc \rectangled{\cmark}\xspace}}
\providecommand{\execfail}{{\sc \rectangled{\xmark}\xspace}}
\providecommand{\execexhaust}{{\sc \rectangled{H}\xspace}}
\providecommand{\txinc}{{\sc \rectangled{I}\xspace}}
\providecommand{\txex}{{\sc \rectangled{X}\xspace}}
\providecommand{\txrevert}{{\sc \rectangled{R}\xspace}}

\begin{abstract}
Transaction simulation is an important subsystem of block building, denial of whose service could lead to severe damage to the blockchain ecosystem and transaction delivery. Denial of block building has been studied, where the existing attack designs either target single-round builders, such as ConditionalExhaust~\cite{DBLP:conf/uss/YaishQZZG24}, or target two-round builders, by exploiting cross-round inconsistency, such as GhostTX~\cite{DBLP:conf/uss/YaishQZZG24} and denial of sequencers~\cite{DBLP:conf/ccs/0001SHC0L0Z25}.

This work examines the denial-of-service security of multi-round transaction simulation under a new exploit: inter-transaction dependency that manifests in smart-contract state.
\end{abstract}


On public blockchains, block building is the process of assembling candidate blocks from unconfirmed transactions, from which downstream validators select the next block to append to the chain. In practice, block building can handle either public or private transactions, operate on-chain or off-chain, and run as a module within validator nodes or as standalone nodes.
Internally, a block-building service's core responsibility is to select unconfirmed transactions from mempools, order and execute them before package into the block to be built. 

%user felt damage!
\noindent{\bf Builder DoS security}:
Block builders control transaction admission in blockchain systems, so denial of their service can significantly disrupt the ecosystem. 
The failure of a builder's service excludes its customers' transactions from the blockchain, hurts the fairness of the upstream MEV-searcher market, centralizes the builder market, and forces downstream relays and proposers to skip victim builders' blocks. In practice, such systematic unfairness can benefit competing builders and MEV searchers and incentivize them to mount the attacks.

%The failure of a single builder reduces its revenue and prevents its customers' transactions from being included, while the simultaneous failure of all builders would cause catastrophic disruption by cutting all transaction senders off from the blockchain. 

Next, we review existing attacks and outline the latest multi-round builder design before introducing our research problem.

\begin{itemize}[leftmargin=*]
\item
\noindent{\bf Existing attacks}:
Denial-of-service attack against block building process has been examined in prior work through several attack vectors, such as disrupting builders' internal mempools~\cite{DBLP:conf/ccs/LiWT21,DBLP:conf/uss/WangT0DY24,DBLP:conf/uss/YaishQZZG24} or exhausting computing resources by transaction flooding~\cite{DBLP:conf/ndss/LiCLT0L21,Tsuchiya_2025,DBLP:conf/ndss/HeoWYKS23} or by executing resource-intensive smart contracts.
The last vector can succeed by evading the protective gas mechanism enforced during block validation~\cite{DBLP:conf/ndss/0002L20,10.1145/3650212.3680372,DBLP:journals/corr/abs-2406-10687} or by exploiting the speculative contract execution prior to validation~\cite{DBLP:conf/uss/YaishQZZG24,DBLP:conf/ndss/LiCLT0L21}.

Among these vectors, speculative contract execution is a promising attack vector. The existing attacks, including ConditionalExhaust~\cite{DBLP:conf/uss/YaishQZZG24}, GhostTX~\cite{DBLP:conf/uss/YaishQZZG24}, denial of sequencers~\cite{DBLP:conf/ccs/0001SHC0L0Z25}, and DoERS~\cite{DBLP:conf/ndss/LiCLT0L21}, exploit factors such as builders' censorship compliance, or RPC contract-testing APIs (e.g., \texttt{eth\_call}), or the inconsistency between the checks inside block building and outside (e.g., the downstream relays in PBS~\cite{DBLP:conf/uss/YaishQZZG24} or the upstream pre-mempool validation~\cite{DBLP:conf/ccs/0001SHC0L0Z25}).

\item
\noindent{\bf Multi-round builders}:
A multi-round builder runs unconfirmed transactions across multiple rounds, each under a distinct transaction ordering. For bundling services, multi-round designs help search for bundles that maximize proposer revenue. The extra rounds also allow filtering out transactions that revert.

\end{itemize}

\noindent{\bf
Attack designs}:
The existing attacks surveyed above are largely agnostic to the internal design of block building and are ineffective against multi-round builders. In brief, their common attack strategy is to passively inspect whether the current execution is speculative and to release the resource-exhaustion payload accordingly. For example, ConditionalExhaust~\cite{DBLP:conf/uss/YaishQZZG24} checks whether the underlying node is censorship-compliant and triggers resource exhaustion only on such nodes. These passive checks, however, cannot distinguish which block-building round a transaction is executing in. Consequently, the resource-exhaustion payloads are triggered prematurely during pre-rounds, where any induced slowdown is confined to adversarial threads and does not interfere with normal transactions executing in parallel threads.

Observing the limitation, we propose a proactive design for evasive attacks on multi-round builders. The key idea is by exploiting {\it inter-transaction dependencies}: On an $n$-round builder, our adversarial contract conceals the resource-exhaustive payload in a path guarded by $(n-1)$ branch conditions, each depending on a distinct state variable. An attack issues more than $n$ transactions, where the first $(n-1)$ set up the context and the remaining transactions continuously trigger the payload. 
As a result, adversarial transactions behave normally in all pre-rounds, evading invalidation, and deliver the payload only in the final round, where adversarial and normal transactions execute together in serial order and delaying one transaction directly delays those that follow.
% This design ensures that only when more than $n$ adversarial transactions run in the same thread, as in the final round, is the exhaustion payload executed, while pre-rounds that split transactions into smaller parallel groups do not trigger it.

Our attack design uses a sequence of transactions that access shared smart contract state. We stress that this exploitation of inter-transaction dependencies has not been considered in prior DoS attack designs. In particular, the most relevant attacks, such as ConditionalExhaust and GhostTX~\cite{DBLP:conf/uss/YaishQZZG24}, use sequences of independent transactions without shared state. The only attack based on the design of inter-dependent transactions is latent overdraft (e.g., DETER-Z in ~\cite{DBLP:conf/ccs/LiWT21}), which is used against mempool, not transaction execution as in this work.

\noindent{\bf 
Mitigation}: We explore the design space of defenses, propose several mitigation strategies, and identify a fundamental challenge in securing any finite-round builder against DoS attacks. 
%Due to space limit, we defer the implementation and evaluation of our defense to Appendix~\ref{sec:mitigate:impl}.

\ignore{\color{blue}
In summary, this work is distinct to related works, notably Yaish et al.~\cite{DBLP:conf/uss/YaishQZZG24}, in three aspects: 1) This work targets multi-round builders with bundling and zk-rollups, while Yaish et al.~\cite{DBLP:conf/uss/YaishQZZG24} studies censorship resistance in mostly single-round builders.
2) The attacks exploit inter-transaction dependencies and varying execution contexts across rounds, making the same transaction behave differently in different rounds, which ~\cite{DBLP:conf/uss/YaishQZZG24} does not use.
3) This work reveals the fundamental hardness in fixing any finite-round design, as they can be evaded with contracts that have more branches; this work calls for future research on alternative round designs that soundly cover all possible transaction contexts before block building.

* The attacks are inherently more sophisticated (for example, using read–write variables in branch conditions) because this complexity is needed to expose these multi-round flaws.
}

\noindent{\bf Summary}: 
This work addresses an open research problem: denial of service against multi-round transaction simulation without relying on cross-round inconsistency. It presents a new attack vector that exploits inter-transaction dependency across rounds in block building.

\bibliographystyle{IEEEtran}
\bibliography{bkc,yuzhetang}

@INPROCEEDINGS {DBLP:journals/corr/abs-2406-10687,
author = { He, Zheyuan and Li, Zihao and Qiao, Ao and Luo, Xiapu and Zhang, Xiaosong and Chen, Ting and Song, Shuwei and Liu, Dijun and Niu, Weina },
booktitle = { 2024 IEEE Symposium on Security and Privacy (SP) },
title = {{ Nurgle: Exacerbating Resource Consumption in Blockchain State Storage via MPT Manipulation }},
year = {2024},
volume = {},
ISSN = {},
pages = {2180-2197},
doi = {10.1109/SP54263.2024.00125},
url = {https://doi.ieeecomputersociety.org/10.1109/SP54263.2024.00125},
publisher = {IEEE Computer Society},
address = {Los Alamitos, CA, USA},
month =May}

@inproceedings{DBLP:conf/ndss/0002L20,
  author    = {Daniel P{\'{e}}rez and
               Benjamin Livshits},
  title     = {Broken Metre: Attacking Resource Metering in {EVM}},
  booktitle = {27th Annual Network and Distributed System Security Symposium, {NDSS}
               2020, San Diego, California, USA, February 23-26, 2020},
  publisher = {The Internet Society},
  year      = {2020},
  url       = {https://www.ndss-symposium.org/ndss-paper/broken-metre-attacking-resource-metering-in-evm/},
  bibsource = {dblp computer science bibliography, https://dblp.org}
}

@inproceedings{DBLP:conf/ndss/HeoWYKS23,
  author       = {Hwanjo Heo and
                  Seungwon Woo and
                  Taeung Yoon and
                  Min Suk Kang and
                  Seungwon Shin},
  title        = {Partitioning Ethereum without Eclipsing It},
  booktitle    = {30th Annual Network and Distributed System Security Symposium, {NDSS}
                  2023, San Diego, California, USA, February 27 - March 3, 2023},
  publisher    = {The Internet Society},
  year         = {2023},
  url          = {https://www.ndss-symposium.org/ndss-paper/partitioning-ethereum-without-eclipsing-it/},
  bibsource    = {dblp computer science bibliography, https://dblp.org}
}

@inproceedings{DBLP:conf/uss/YaishQZZG24,
  author       = {Aviv Yaish and
                  Kaihua Qin and
                  Liyi Zhou and
                  Aviv Zohar and
                  Arthur Gervais},
  editor       = {Davide Balzarotti and
                  Wenyuan Xu},
  title        = {Speculative Denial-of-Service Attacks In Ethereum},
  booktitle    = {33rd {USENIX} Security Symposium, {USENIX} Security 2024, Philadelphia,
                  PA, USA, August 14-16, 2024},
  publisher    = {{USENIX} Association},
  year         = {2024},
  url          = {https://www.usenix.org/conference/usenixsecurity24/presentation/yaish},
  bibsource    = {dblp computer science bibliography, https://dblp.org}
}

@inproceedings{10.1145/3650212.3680372,
author = {Wu, Shuohan and Li, Zihao and Zhou, Hao and Luo, Xiapu and Li, Jianfeng and Wang, Haoyu},
title = {Following the “Thread”: Toward Finding Manipulatable Bottlenecks in Blockchain Clients},
year = {2024},
isbn = {9798400706127},
publisher = {Association for Computing Machinery},
address = {New York, NY, USA},
url = {https://doi.org/10.1145/3650212.3680372},
doi = {10.1145/3650212.3680372},
booktitle = {Proceedings of the 33rd ACM SIGSOFT International Symposium on Software Testing and Analysis},
pages = {1440–1452},
numpages = {13},
location = {Vienna, Austria},
series = {ISSTA 2024}
}

@article{Tsuchiya_2025,
   title={Blockchain Amplification Attack},
   volume={9},
   ISSN={2476-1249},
   url={http://dx.doi.org/10.1145/3711697},
   DOI={10.1145/3711697},
   number={1},
   journal={Proceedings of the ACM on Measurement and Analysis of Computing Systems},
   publisher={Association for Computing Machinery (ACM)},
   author={Tsuchiya, Taro and Zhou, Liyi and Qin, Kaihua and Gervais, Arthur and Christin, Nicolas},
   year={2025},
   month=mar, pages={1–30} }

@inproceedings{DBLP:conf/ccs/0001SHC0L0Z25,
  author       = {Zihao Li and
                  Zhiyuan Sun and
                  Zheyuan He and
                  Jinzhao Chu and
                  Hao Zhou and
                  Xiapu Luo and
                  Ting Chen and
                  Yinqian Zhang},
  editor       = {Chun{-}Ying Huang and
                  Jyh{-}Cheng Chen and
                  Shiuh{-}Pyng Shieh and
                  David Lie and
                  V{\'{e}}ronique Cortier},
  title        = {Denial of Sequencing Attacks in Ethereum Layer 2 Rollups},
  booktitle    = {Proceedings of the 2025 {ACM} {SIGSAC} Conference on Computer and
                  Communications Security, {CCS} 2025, Taipei, Taiwan, October 13-17,
                  2025},
  pages        = {2084--2098},
  publisher    = {{ACM}},
  year         = {2025},
  url          = {https://doi.org/10.1145/3719027.3765100},
  doi          = {10.1145/3719027.3765100},
  bibsource    = {dblp computer science bibliography, https://dblp.org}
}

@inproceedings{DBLP:conf/uss/WangT0DY24,
  author       = {Yibo Wang and
                  Yuzhe Tang and
                  Kai Li and
                  Wanning Ding and
                  Zhihua Yang},
  editor       = {Davide Balzarotti and
                  Wenyuan Xu},
  title        = {Understanding Ethereum Mempool Security under Asymmetric DoS by Symbolized
                  Stateful Fuzzing},
  booktitle    = {33rd {USENIX} Security Symposium, {USENIX} Security 2024, Philadelphia,
                  PA, USA, August 14-16, 2024},
  publisher    = {{USENIX} Association},
  year         = {2024},
  url          = {https://www.usenix.org/conference/usenixsecurity24/presentation/wang-yibo},
  bibsource    = {dblp computer science bibliography, https://dblp.org}
}

@inproceedings{DBLP:conf/ndss/LiCLT0L21,
  author    = {Kai Li and
               Jiaqi Chen and
               Xianghong Liu and
               Yuzhe Richard Tang and
               XiaoFeng Wang and
               Xiapu Luo},
  title     = {As Strong As Its Weakest Link: How to Break Blockchain DApps at {RPC}
               Service},
  booktitle = {28th Annual Network and Distributed System Security Symposium, {NDSS}
               2021, virtually, February 21-25, 2021},
  publisher = {The Internet Society},
  year      = {2021},
  url       = {https://www.ndss-symposium.org/ndss-paper/as-strong-as-its-weakest-link-how-to-break-blockchain-dapps-at-rpc-service/},
  bibsource = {dblp computer science bibliography, https://dblp.org}
}

@inproceedings{DBLP:conf/ccs/LiWT21,
  author    = {Kai Li and
               Yibo Wang and
               Yuzhe Tang},
  editor    = {Yongdae Kim and
               Jong Kim and
               Giovanni Vigna and
               Elaine Shi},
  title     = {{DETER:} Denial of Ethereum Txpool sERvices},
  booktitle = {{CCS} '21: 2021 {ACM} {SIGSAC} Conference on Computer and Communications
               Security, Virtual Event, Republic of Korea, November 15 - 19, 2021},
  pages     = {1645--1667},
  publisher = {{ACM}},
  year      = {2021},
  url       = {https://doi.org/10.1145/3460120.3485369},
  doi       = {10.1145/3460120.3485369},
  bibsource = {dblp computer science bibliography, https://dblp.org}
}

\end{document}